\documentclass[5p,twocolumn,times,number]{elsarticle}
\usepackage{amsmath}
\usepackage{amssymb}
\usepackage{graphics}
\usepackage{graphicx}
\usepackage{epsfig}
\usepackage{dcolumn}
\usepackage{bm}
\begin{document}
\begin{frontmatter}

\title{Systematic measurements of the gain and the energy resolution of single and double mask GEM detectors}
\author[label1,label2]{S.~Biswas\corref{cor}}
\ead{saikat.ino@gmail.com, s.biswas@niser.ac.in, saikat.biswas@cern.ch}
%\thanks[label1]{}
%\corauth[cor1]{}
\author[label1]{D.~J.~Schmidt}
\author[label1,label4]{A.~Abuhoza}
\author[label1]{U.~Frankenfeld}
\author[label1]{C.~Garabatos}
\author[label1]{J.~Hehner}
\author[label1]{V.~Kleipa}
\author[label1]{T.~Morhardt}
\author[label1]{C.~J.~Schmidt}
\author[label3]{H.~R.~Schmidt}
\author[label3]{J.~Wiechula}

\cortext[cor]{Corresponding author}

\address[label1]{GSI Helmholtzzentrum f\"ur Schwerionenforschung GmbH, Planckstrasse 1, D-64291 Darmstadt, Germany}
\address[label2]{School of Physical Sciences, National Institute of Science Education and Research, Jatni - 752050, India}
\address[label3]{Eberhard-Karls-Universit\"at, T\"ubingen, Germany}
\address[label4]{King Abdulaziz City forv Science and Technology (KACST), Riyadh, Saudi Arabia}

\begin{abstract}
Systematic studies on the gain and the energy resolution have been carried out varying the voltage across the GEM foils for both single mask and double mask triple GEM detector prototypes. Variation of the gain and the energy resolution have also been measured varying either the drift voltage, transfer voltage and induction voltage keeping other voltages constant. The results of the systematic measurements has been presented.
\end{abstract}
\end{frontmatter}

\section{Introduction}\label{intro}
Triple GEM detectors will be used to instrument the CBM (Compressed Baryonic Matter) muon detector MUCH (MUon CHamber) at the future Facility for Antiproton and Ion Research (FAIR) in Darmstadt, Germany \cite{CBM,FAIR,CBM2008,FS97}. In the GSI detector laboratory an R\&D effort has been performed to study the characteristics of both the single and double mask GEM detectors \cite{CBM11,SB11,SB12}. In this study, the gain and the energy resolution have been measured systematically employing an Fe$^{55}$ source as a function of the voltages applied to the GEM foils and different gaps. In this article the results of the systematic measurements have been presented.

\section{Description of the GEM modules and electronics}\label{construct}
In this measurement both single-mask and  double-mask standard triple GEM detectors of area 10~cm~$\times$~10~cm have been used. The GEM foils are obtained from CERN. The drift gap, 2-transfer gaps and induction gap has been kept of 3,~2,~2,~2~mm respectively for both type of chambers. The read out plane consist of 256 number of 6$\times$6 mm$^2$ pads. All the readout pads are connected to four connectors each with 64 connections. Although there are a segmented readout pad for the chamber, in this study the signals obtained from all the pads from one 64 pin connectors, summed by an sum-up board and a single input has been fed to a charge sensitive preamplifier. The output of the preamplifier has been fed to a PXI LabVIEW based data acquisition system \cite{NI}. In this systematic measurement a premixed gas of Argon and CO$_2$ in 70/30 volume ratio has been used for the chambers.

The high voltage (HV) to the GEM is applied by a seven-channel HVG210 power supply made by LNF-INFN \cite{GC}. The HVG210 module allows for controlling the power supply voltage of a triple GEM detector. The module communicates with peripherals via CAN bus. HVG210 power supply is composed by seven almost identical sections each of them being able to produce a specified voltage level with a current limiting option. 11 M$\Omega$ $(10+1) [R]$ protection resistance have been used on the top of all the GEM foils and on the drift plane. This particular HV module has an advantage that individual voltage to different GEM planes can be applied and current from individual channels can be read out.

\section{Measurements and results}\label{res}
\begin{figure}[htb!]
\begin{center}
\includegraphics[scale=0.34]{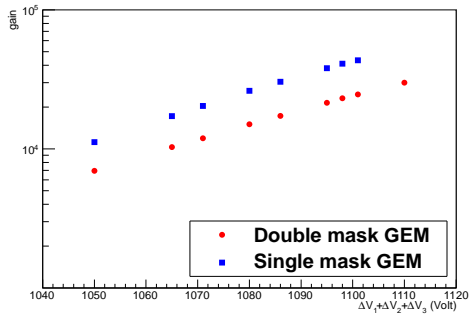}
\caption{Variation of the gain as a function of the GEM global voltage.}\label{gain_spectrum_gem}
\end{center}
\end{figure}

\begin{figure}[htb!]
\begin{center}
\includegraphics[scale=0.34]{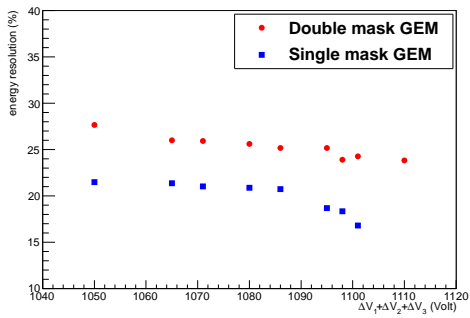}
\caption{Variation of the energy resolution as a function of the GEM global voltage.}\label{reso_gem}
\end{center}
\end{figure}

In this study, the gain of the detectors have been measured by obtaining the mean position of 5.9~keV peak of Fe$^{55}$ X-ray spectrum with Gaussian fitting \cite{SB11}. The \% energy resolution is defined as $\frac{\sigma~\times~2.355}{mean}~\times~100$, where $\sigma$ is the standard deviation of the Gaussian peak. During these measurements the drift field, transfer fields and the induction field are kept constant at 2.33~kV/cm, 3.25~kV/cm and 3.25~kV/cm respectively for both the single mask and double mask GEM detectors. The variation of the gain and energy resolution of these detectors with that of the global GEM voltage ($\Delta$V$_{1}$+$\Delta$V$_{2}$+$\Delta$V$_{3}$) are shown in Figure~\ref{gain_spectrum_gem} and Figure~\ref{reso_gem} respectively. For both the detectors the gain increases exponentially with the global GEM voltage. The energy resolution value decreases as gain increases for both the chambers.
%%%%%%%%%%%%%%%%%%%%%%%%%%%%%%%%%%%%%%%%%%%%%%%%%%%%%%%%%%%%%%%%%%%
%%%%%%%%%%%%%%%%%%%%%%%%%%%%%%%%%%%%%%%%%%%%%%%%%%%%%%%%%%%%%%%%%%%
%%%%%%%%%%%%%%%%%%%%%%%%%%%%%%%%%%%%%%%%%%%%%%%%%%%%%%%%%%%%%%%%%%%
%%%%%%%%%%%%%%%%%%%%%%%%%%%%%%%%%%%%%%%%%%%%%%%%%%%%%%%%%%%%%%%%%%%

\begin{figure}[htb!]
\begin{center}
\includegraphics[scale=0.34]{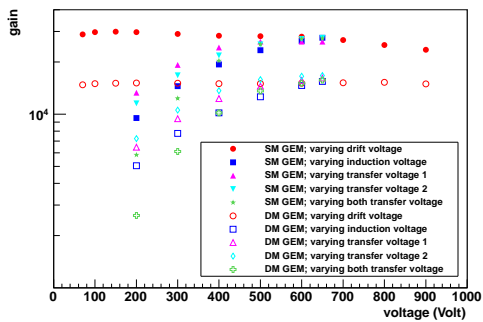}
\caption{Variation of the gain as a function of the different voltages.}\label{gain_spectrum}
\end{center}
\end{figure}

\begin{figure}[htb!]
\begin{center}
\includegraphics[scale=0.34]{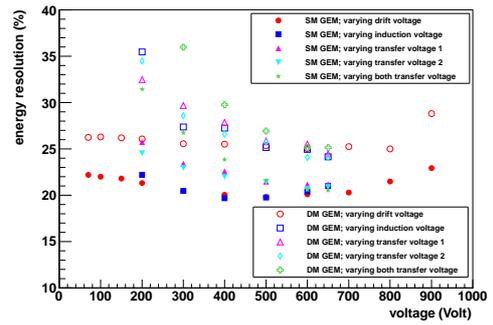}
\caption{Variation of the energy resolution as a function of the different voltages.}\label{reso}
\end{center}
\end{figure}

The variation of the gain and the energy resolution have also been measured varying the drift voltage, induction voltage, transfer voltage 1, transfer voltage 2 and both the transfer voltage simultaneously. During these all measurements the voltage setting across the GEMs are kept at $\Delta V_1$=365~V, $\Delta V_2$=360~V and $\Delta V_3$=355~V respectively from top (drift plane side) to bottom (pad plane side). For each set of readings, say during the drift voltage variation all other voltages such as induction voltage and transfer voltages are kept constant according to the corresponding electric fields mentioned above. The variation of the gain and the energy resolution of both these detectors as a function of the respective voltage variation are shown in Figure~\ref{gain_spectrum} and Figure~\ref{reso} respectively.
%%%%%%%%%%%%%%%%%%%%%%%%%%%%%%%%%%%%%%%%%%%%%%%%%%%%%%%%%%%%%%%%%%%
%%%%%%%%%%%%%%%%%%%%%%%%%%%%%%%%%%%%%%%%%%%%%%%%%%%%%%%%%%%%%%%%%%%
%%%%%%%%%%%%%%%%%%%%%%%%%%%%%%%%%%%%%%%%%%%%%%%%%%%%%%%%%%%%%%%%%%%
%%%%%%%%%%%%%%%%%%%%%%%%%%%%%%%%%%%%%%%%%%%%%%%%%%%%%%%%%%%%%%%%%%%

\section{Conclusions}
In the GSI detector laboratory a systematic study has been carried out to measure the variation the gain and the energy resolution as a function of the voltages applied to the GEM foils and different gaps such as drift gap, transfer gaps and the induction gap, employing an Fe$^{55}$ source for both the single and double mask GEM detectors. It has been observed that for both the detectors gain increases exponentially with the global GEM voltage. It has also been observed that for very low and very high drift voltage the gain is somewhat reduced, while it is nearly constant at the intermediate values. In case of other voltage variations such as the induction and transfer voltages, it has been observed that the gain increases with the voltage and saturates at some point. The energy resolution always decreases as gain increases.

\section{Acknowledgements}
We are thankful to Dr. Ingo Fr\"{o}hlich of University of Frankfurt, Prof. Dr. Peter Fischer of Institut f\"{u}r Technische Informatik der Universit\"{a}t Heidelberg, Prof. Dr. Peter Senger, CBM Spokesperson and Dr. Subhasis Chattopadhyay, Deputy spokesperson, CBM  for their support in course of this work. S. Biswas acknowledges the support of DST-SERB Ramanujan Fellowship (D.O. No. SR/S2/RJN-02/2012).

\end{document}